\documentstyle[12pt]{article}
\begin{document}
\begin{flushright}
UT-Komaba 97-13
\end{flushright}
\begin{center} 
{\Large{\bf  Quasi-Excitations and Superconductivity \\ in the t-J model
on a Ladder}}
\vskip 0.5cm
 {\Large  Ikuo Ichinose$^{\star}$\footnote{e-mail address: 
ikuo@hep1.c.u-tokyo.ac.jp} and
  Tetsuo Matsui$^{\dagger}$ \footnote{e-mail address: 
matsui@phys.kindai.ac.jp}}
 \vskip 0.1in
 $^{\star}$Institute of Physics, University of Tokyo, Komaba, Tokyo, 153 Japan  \\
 $^{\dagger}$Department of Physics, Kinki University, Higashi-Osaka,  577 Japan 
\end{center}

\begin{center} 
\begin{bf}
Abstract
\end{bf}
\end{center}

We study the t-J model on a ladder using the slave-fermion-CP$^1$
formalism which has been used successfully in studying 
lightly doped high-T$_C$ cuprates.
Special attention is paid to the dynamics of composite gauge fields and
the  natures of quasi-excitations. 
The slave-fermion-CP$^1$ approach explains many aspects of the ladder
system observed
by experiments and numerical studies in a  natural and coherent manner.
We first obtain the low-energy effective model by integrating 
out half of the CP$^1$
variables (the Schwinger bosons) assuming a {\it short-range }
antiferromagnetic order (SRAFO).
The  spin part of the effective model is the relativistic CP$^1$ model.
In the single-chain case, there appears a topological $\theta$-term
with $\theta=\pi$, as is well known.
On the other hand, in the two-leg ladder case, we have $\theta=2\pi$.
The dynamics of the composite gauge boson strongly depends on the value
of the coefficient of the $\theta$-term.
This fact explains why the quasi-excitations in a chain and 
ladders of even legs are different. For a ladder, the gauge dynamics   
realizes in the confinement phase, so the quasi-exciations are 
charge-neutral objects like spin triplet and electrons, etc.
The effective model reveals attractive force between holes,
which generates  superconductivity in SRAFO.
The symmetry of hole-pair condensation should be of the d-wave type.

\vfill
\eject

\setcounter{footnote}{0}
\section{Introduction}
In the last few years, quasi-one-dimensional systems of strongly
correlated electrons are of great theoretical and experimental 
interest in condensed matter physics\cite{DR}.
Studies on these systems are expected to shed light on the mechanism 
of high-T$_C$ superconductivity.
At present, it is known that spin excitations in an  
antiferromagnetic (AF) Heisenberg model put on an even-leg ladder have an 
energy gap, though those on an odd-leg 
ladder are gapless.
Very recently, superconductivity has been observed in 
(weakly interacting) doped spin ladders with a spin gap\cite{supex},
and the  symmetry of the  order parameter seems to be of a d-wave 
type as some theoretical studies predict\cite{supth}.

The metallic phase of high-T$_C$ cuprates is anormalous, and a phenomenon
called charge-spin separation (CSS) is expected to occur\cite{Anderson}.
In  mean-field theories (MFT) of the t-J model 
in  slave-particle representations,
the CSS is implemented naturally.
However, the  phase degrees of freedom of  MF's behave like 
(composite) gauge fields\cite{gauge}, so 
careful study is necessary in order to check the reliability of MFT.
In the previous papers\cite{IMCSS1,IMCSS2}, we argued 
 that  the CSS takes place at sufficiently low temperatures 
$(T)$ in the two-dimensional (2D) t-J model.
To derive the result, a gauge-theoretical  treatment was essential.

In the present paper, we shall study the t-J model on a ladder of two legs
by using the slave-fermion (SF) representation.
The  gauge-theoretical method is also applicable for this model
straightfowardly. 
It clarifies
why the quasi-excitations in  chain and ladder systems are different
so much; In a chain, 
the full CSS takes place and holons and spinons are 
quasi-excitations, while 
in a two-leg ladder, we shall see that the system is in a confinement phase and 
the gauge-charge-neutral
objects like spin triplet and bound states of holons and 
spinons are low-energy excitations.
This result explains the numerical studies in Ref.\cite{TTR}.

There are field-theoretical studies for the ladder systems,
 most of which use the bosonization techniques\cite{boson}.
We want to stress here that the gauge-theoretical study is useful 
not only for the 2D systems of strongly-correlated electrons but also 
for the  quasi-one-dimensional systems.
 Furthermore, we believe that it gives us an universal and coherent
 understanding of a wide variety of strongly-correlated electron
systems including the fractional quantum Hall effect.

This paper is organized as follows.
In Sect.2, we shall introduce the t-J model in the SF-CP$^1$
representation.
This procedure is closely related with our previous work which studied
the 2D t-J model\cite{IMSF}.
In Sect.3, we shall obtain a low-energy effective model by integrating out
half of the spin variables (CP$^1$ variables) 
assuming a short-range AF order (SRAFO).
The effective model, as  a result of the existence of SRAFO,
contains important interactions among  
smooth spin variables and hole field.
In Sect.4, we shall study the spin dynamics in the chain and 
ladder t-J model.
The continuum limit of the spin part of the effective model is the CP$^1$
model with a topological term.
Coefficient of the topological term depends on the number of legs 
and strongly influences the dynamics of composite gauge bosons.
As we explained previously\cite{IMCSS1,IMCSS2}, 
the mechanism of  CSS in  strongly-correlated electron systems 
is described by a (de)confinement phenomenon
of composite gauge fields.
In Sect.5 we shall study the dynamics of composite gauge bosons 
on a two-leg ladder using the method presented in \cite{IMCSS2}
and identify quasi-excitations. We find that the system
is in the confinement phase down to $T$ = 0.
So the quasi-exciations should be gauge-charge-neutral bound states
of holons and spinons, like a spin triplet (magnons) and electrons. 
In Sect.6, we shall study the possibility of superconductivity 
and its symmetry.
The effective model involves an attractive force between holes in the
SRAFO background which enhances a d-wave hole-pair condensation.
In Sect.7 we present discussion.

\section{The t-J model in the SF-CP$^1$ representation}
\setcounter{equation}{0}

We consider the t-J model on a two-leg ladder whose Hamiltonian 
is given by
\begin{eqnarray}
H&=&-\sum_{i,\sigma}\Big( t \sum^2_{a=1} 
C^{\dagger}_{i+1,a,\sigma}C_{i,a,\sigma}
+t'C^{\dagger}_{i,1,\sigma}C_{i,2,\sigma}+H.c.\Big)  \nonumber   \\
&+&\sum_i \Big[ J\sum^2_{a=1}\Big(\vec{S}_{i+1,a}\vec{S}_{i,a}-
{1 \over 4}n_{i+1,a}n_{i,a}\Big) +J'\Big(\vec{S}_{i,1}\vec{S}_{i,2}-{1\over 4}n_{i,1}n_{i,2}\Big)\Big].
\label{HtJ}
\end{eqnarray}
Here the suffix $a (=1,2)$ distinguishes one of  two legs, $i$ labels 
sites along each leg.  $C_{i,a,\sigma}$
is the electron operator with the spin $\sigma ( = 1,2)$, and 
$\vec{S}_{i,a}$ and $n_{i,a}$ are the spin and number operators 
on the site $(i,a)$. The  physical states must satisfy 
\begin{equation}
n_{i,a} <2,
\label{physcon}
\end{equation}
on each site.
In the SF formalism, the electron operator is expressed in terms of
the bosonic spinon operators $a_{i,a,\sigma}$ and the fermionic 
holon $\psi_{i,a}$ operators as
\begin{equation}
C_{i,a,\sigma}=\psi^{\dagger}_{i,a}a_{i,a,\sigma},
\label{slaveF}
\end{equation}
and the physical state condition (\ref{physcon}) becomes
\begin{equation}
\sum_{\sigma} a^{\dagger}_{i,a,\sigma}a_{i,a,\sigma}+\psi^{\dagger}_{i,a}
\psi_{i,a}=1.
\label{physcon2}
\end{equation}
As in the previous paper \cite{IMSF}, we solve
the condition (\ref{physcon2}) by rewriting the operator
$a_{i,a}$  in terms of CP$^1$ operator $z_{i,a}$ as
\begin{eqnarray}
a_{i,a,\sigma} &=& (1-\psi^{\dagger}_{i,a}\psi_{i,a})^{1/2}z_{i,a,\sigma}  
\nonumber  \\
&=& (1-\psi^{\dagger}_{i,a}\psi_{i,a})z_{i,a,\sigma},  
\label{CP1}
\end{eqnarray}
\begin{equation}
\sum_{\sigma}  z^{\dagger}_{i,a,\sigma}z_{i,a,\sigma}=1.
\label{CP1v}
\end{equation}
This representation is quite useful especially for the lightly doped case
of the t-J model and the AF Heisenberg model\cite{IMSF,YTIM}.
We shall treat $\psi_{i,a}$ and $z_{i,a,\sigma}$ as fundamental variables
and employ the path-integral formalism (See Ref.\cite{IMSF} 
for detailed discussions on this formalism). 

From (\ref{slaveF}) and (\ref{CP1}), it is obvious that there appears
a local gauge symmetry in the SF-CP$^1$ t-J model;
\begin{eqnarray}
z_{i,a,\sigma} &\rightarrow & e^{i\phi_{i,a}}z_{i,a,\sigma},  \nonumber  \\
\psi_{i,a} &\rightarrow & e^{i\phi_{i,a}}\psi_{i,a}.
\label{gaugetrf}
\end{eqnarray}
The electron operator $C_{i,a,\sigma}$ is invariant under this 
transformation since it is a composite of a spinon and an anti-holon.
One may expect that there appears composite gauge bosons with
respect to (\ref{gaugetrf}).
As we showed in the previous papers, these gauge bosons are introduced
as auxiliary fields in path-integrals, and their dynamics specifies
the nature of quasi-excitations.

\section{Short-range AF and low-energy effective model}
\setcounter{equation}{0}

In this section we shall obtain the low-energy effective model 
by integrationg out half of the CP$^1$ variables $z_{i,a,\sigma}$ 
(e.g., the CP$^1$ variables on the 
odd sites) assuming a SRAFO.

In the path-integral formalism, the  partition function $Z$ is given as 
\begin{eqnarray}
Z &=& \int [Dz][D\psi] \exp \Big[ \int d\tau A(\tau) \Big],  \nonumber  \\
A(\tau) &=& -\sum_{i,a}
\Big(\sum_{\sigma}\bar{z}_{i,a,\sigma}\dot{z}_{i,a,\sigma}
+\bar{\psi}_{i,a}\dot{\psi}_{i,a}\Big)-H.
\label{partition}
\end{eqnarray}
The J-terms in $H$ is explicitly given in terms of  $z_{i,a,\sigma}$
and $\psi_{i,a}$ as
\begin{eqnarray}
A_J &=& -{J \over 2} \sum_{i,a}\rho^2_{i,a}\rho^2_{i+1,a}[(\bar{z}_{i,a}
z_{i+1,a})(\bar{z}_{i+1,a}z_{i,a})-1],   \nonumber  \\
A_{J'} &=& -{J' \over 2}\sum_i \rho^2_{i,1}\rho^2_{i,2}
[(\bar{z}_{i,1}z_{i,2})(\bar{z}_{i,2}z_{i,1})-1],
\label{AJ}
\end{eqnarray}
where $\rho^2_{i,a}=1-\bar{\psi}_{i,a}\psi_{i,a}$. 

From (\ref{AJ}), it is obvious that the SRAFO configurations,  
$\vec{S_j} \simeq -\vec{S_i}$ for a nearest-neighbor (NN) pair $(i,j)$, give
dominant contributions to the path integral. One can express
a spin $\vec{S_{i}}$ and its time-reversed spin $-\vec{S_{i}}$ as
\begin{eqnarray}
\vec{S_{i}} & = & \bar{z_{i}}\vec{\sigma} z_{i} \nonumber\\
-\vec{S_{i}} & = &  \bar{\tilde{z}}_{i}\vec{\sigma} \tilde{z}_{i} \nonumber\\
\tilde{z}_{\sigma}&\equiv & \epsilon_{\sigma \sigma'}\bar{z}_{\sigma'},
\; \; \epsilon_{12} = -\epsilon_{21} = 1,
\end{eqnarray} 
where  $\vec{\sigma}$ is the Pauli spin matrices.
Therefore the assumption of SRAFO is written as 
\begin{equation}
z_i \simeq  \tilde{z}_j, \; \;  
\langle i,j \rangle =\mbox{NN sites}.
\label{SRAF}
\end{equation} 

To integrate out half of the CP$^1$ variables, say, those on the odd sites,
around the SRAF configurations (\ref{SRAF}),
we pair every odd site with a NN even site and denote the CP$^1$ variables
in a  pair as $z_o$ and $z_e$.
For $J \geq J'$, we take $z_e=z_{i+1,a}$ for odd-site $z_o=z_{i,a}$,
whereas for $J' \geq J$, $z_e=z_{i,\bar{a}}$ for $z_o=z_{i,a}$ where 
$\bar{a}=2$ for $a=1$ and $\bar{a}=1$ for $a=2$ (see Fig.1). 

The odd site CP$^1$ variable $z_o$ is parameterized as follows 
by using its reference coordinate $z_e$,
\begin{equation}
z_o=p_{oe}z_e+q_{oe}\tilde{z}_e, \;\; 
q_{oe}=(1-\bar{p}_{oe}p_{oe})^{{1\over 2}}
\; U_{oe}, \;\; U_{oe} \in U(1).
\label{zpara}
\end{equation}
We substitute (\ref{zpara}) into $A(\tau)$ and expand 
it in powers of $p_{oe}$ up to $O(p_{oe}^2)$;
\begin{equation}
A=A_0+A_p+O(p^3_{oe}), \nonumber
\end{equation}
\begin{eqnarray}
A_0 &=& \sum_{o\in odd}\Big[ -\bar{\psi}_e\dot{\psi}_e
-\bar{\psi}_o\dot{\psi}_o
+(\rho^2_o-\rho^2_e)\bar{z}\dot{z}+\mu_c(\rho^2_o+\rho^2_e)  \nonumber  \\
&& +{1 \over 2}\rho^2_o 
\sum_{\nu=u,d,s}J_{\nu}\rho^2_{o\nu}(\bar{z}z_{o\nu})
(\bar{z}_{o\nu}z)-\sum_{\nu=u,d,s} t_{\nu}\rho_o\rho_{o\nu}[\bar{\psi}_o
\psi_{o\nu}U_{oe}(\bar{z}_{o\nu}\tilde{z})  \nonumber  \\
&& +\bar{U}_{oe}(\bar{\tilde{z}}z_{o\nu})\bar{\psi}_{o\nu}\psi_o]\Big],
\label{A0}
\end{eqnarray}
where the suffix $\nu ( u,d,s)$ denotes the three NN directions from an
odd site $o$, and we omit the suffix $e$ as $z=z_e$.
 $\mu_c$ is the chemical potential to enforce the hole concentration 
to be $\langle \bar{\psi}_i \psi_i \rangle = \delta$. $J_{\nu}, t_{\nu}$
denote
\begin{equation}
J_{\nu}=\left\{
               \begin{array}{ll}
	       J, & \quad \mbox{$\nu$=u,d}   \\
	       J', & \quad \mbox{$\nu$=s}
	       \end{array} \right . \label{Jnu}
\end{equation}	       	   
\begin{equation}
t_{\nu}=\left\{
               \begin{array}{ll}
	       t, & \quad \mbox{$\nu$=u,d}   \\
	       t', & \quad \mbox{$\nu$=s}.
	       \end{array} \right . \label{tnu1}
\end{equation}	       	   

$A_p$ is given by
\begin{equation}
A_p=\sum_{o\in odd}(-\bar{p}_{oe}M_op_{oe}+\bar{p}_{oe}k_{oe}
+\bar{l}_{oe}p_{oe}),
\label{Ap}
\end{equation}
\begin{eqnarray}
M_o &=& {1\over 2}\rho^2_o\overrightarrow{\partial_{\tau}}-{1\over 2}
\overleftarrow{\partial_{\tau}}\rho^2_o+2\rho^2_o\bar{z}\dot{z}  \nonumber   \\
&& +{1\over 2}\sum_{\nu}J_{\nu}\rho^2_o\rho^2_{o\nu}
[(\bar{z}_{o\nu}z)(\bar{z}z_{o\nu})-(\bar{z}_{o\nu}\tilde{z})(\bar{\tilde{z}}z_{o\nu})] \nonumber  \\
&& -{1 \over 2}\sum_{\nu} t_{\nu}\rho_o\rho_{o\nu}[\bar{\psi}_o\psi_{o\nu}U_{oe}
(\bar{z}_{o\nu}\tilde{z})+\bar{U}_{oe}(\bar{\tilde{z}}z_{o\nu})\bar{\psi}_{o\nu}\psi_o],  \label{Mo} \\
k_{oe} &=& -\rho^2_oU_{oe}\bar{z}\dot{\tilde{z}}-\sum_{\nu}\Big[{J_{\nu} \over 2}
\rho^2_o\rho^2_{o\nu}(\bar{z}z_{o\nu})(\bar{z}_{o\nu}\tilde{z})U_{oe}  \nonumber  \\
&& +t_{\nu}\rho_o\rho_{o\nu}\bar{\psi}_{o\nu}\psi_o(\bar{z}z_{o\nu})\Big], \nonumber   \\
\bar{l}_{oe} &=& -\rho^2_o\bar{U}_{oe}\bar{\tilde{z}}\dot{z}-\sum_{\nu}
\Big[{J_{\nu} \over 2}\rho^2_o\rho^2_{o\nu}(\bar{\tilde{z}}z_{o\nu})(\bar{z}_{o\nu}z)
\bar{U}_{oe}  \nonumber   \\
&& +t_{\nu}\rho_o\rho_{o\nu}\bar{\psi}_o\psi_{o\nu}(\bar{z}_{o\nu}z)\Big].
\label{kloe}
\end{eqnarray}
We do Gaussian integration over $p_{oe}$'s;
\begin{equation}
\int [dp] \exp \Big[ \int d\tau A_p\Big]=\prod_{o\in odd} (\det M_o)^{-1}\cdot
\exp \Big[\int d\tau A_1(\tau) \Big].
\label{intp}
\end{equation}
At low $T$, $T< 2J+J'$, this reduces to \cite{IMSF}
\begin{eqnarray}
\int d\tau A_1(\tau) &=& \int d\tau d\tau' \sum_o \bar{l}_{oe}(\tau)M^{-1}_o(\tau,\tau')
k_{oe}(\tau')  \nonumber  \\
&\sim & {2 \over 2J+J'} \int d\tau\sum_o\bar{l}_{oe}(\tau)k_{oe}(\tau).
\label{A1}
\end{eqnarray}

From (\ref{A0}) and (\ref{A1}), the relevant terms of $z$ and $\psi$ are readily 
obtained.
For the CP$^1$ field $z$,
\begin{eqnarray}
A^{z}_0 &=& -{1 \over 2} \sum_oJ_{\nu}(\bar{z}\tilde{z}_{o\nu})(\bar{\tilde{z}}_{o\nu}
z),  \nonumber  \\
A^z_1 &=& 2\sum_o (2J+J')^{-1}\Big[ (\bar{\tilde{z}}\dot{z})(\bar{z}\dot{\tilde{z}}) \nonumber  \\
 && \; +\sum_{\nu\nu'}{J_{\nu} \over 2}{J'_{\nu} \over 2}(\bar{\tilde{z}}z_{o\nu})
 (\bar{z}_{o\nu}z)(\bar{z}z_{o\nu'})(\bar{z}_{0\nu'}\tilde{z})  \nonumber   \\
&& \; +(\bar{\tilde{z}}\dot{z})\sum_{\nu}{J_{\nu} \over 2} (\bar{z}z_{o\nu})
(\bar{z}_{o\nu}\tilde{z})+(\bar{z}\dot{\tilde{z}})\sum_{\nu}{J_{\nu} \over 2}
(\bar{\tilde{z}}z_{o\nu})(\bar{z}_{o\nu}z)\Big].
\label{Az}
\end{eqnarray}
Similarly, for the hole field $\psi$, we obtain
\begin{eqnarray}
A^{\psi}_ {rel}&=& K+T_0+T_1+T_2,  \nonumber  \\
K &=& -\sum_o\Big[ \bar{\psi}(D_{\tau}+m)\psi+\bar{\eta}(D_{\tau}-m)\eta\Big],
\;\; m=\mu_c+2J+J',  \nonumber   \\
T_0 &=& -{1\over 2}\sum_o\sum_{\nu}J_{\nu}\bar{\eta}_o\eta_o\bar{\psi}_{o\nu}
\psi_{o\nu}(\bar{z}z_{o\nu})(\bar{z}_{o\nu}z),  \nonumber  \\
T_1 &=& \sum_o\sum_{\nu}t_{\nu}(b_{\nu}\bar{\psi}_{o\nu}\bar{\eta}_{o}
+c_{\nu}\eta_o\psi_{o\nu}),  \nonumber   \\
b_{\nu} &=& -(\bar{\tilde{z}}z_{o\nu})+{(\bar{z}z_{o\nu}) \over 2J+J'}
\Big[2(\bar{\tilde{z}}\dot{z})+\sum_{\nu'}J_{\nu'}(\bar{z}_{o\nu'}z)
(\bar{\tilde{z}}z_{o\nu'})\Big], \nonumber\\
c_{\nu}&=& -(\bar{z}_{o\nu} \tilde{z})+{(\bar{z}_{o\nu }z) \over 2J+J'}
\Big[2(\bar{ z }\dot{\tilde{z}})+\sum_{\nu'}J_{\nu'}(\bar{z}z_{o\nu'})
(\bar{z}_{o\nu'}\tilde{z})\Big],  \nonumber  \\
T_2 &=& {2 \over 2J+J'}\sum_o\sum_{\nu\nu'}t_{\nu}t_{\nu'}\eta_o\psi_{o\nu}
\bar{\psi}_{o\nu'}\bar{\eta}_{o}(\bar{z}_{o\nu}z)(\bar{z}z_{o\nu'}), 
\label{Apsi}
\end{eqnarray} 
where 
\begin{equation}
D_{\tau}=\partial_{\tau}+iA_{\tau}=\partial_{\tau}-(\bar{z}\dot{z}).
\label{Dtau}
\end{equation}
Above, we have defined 
\begin{equation}
\eta_o=U_{oe}\bar{\psi}_o,
\label{eta}
\end{equation}
which transforms $\eta_o \rightarrow e^{i\phi_e}\eta_o$ under (\ref{gaugetrf}).
The action of the effective lattice model is thus given as
$A_{\rm{eff}}=A^z_0+A^z_1+A^{\psi}_{rel}$.

$T_0$ and $T_2$ in the effective model show that there appear effective 
interactions between holes as a result of the SRAFO.
One can expect that a superconducting phase appears when
weak-inter-ladder interactions are included as recently observed 
by experiments\cite{supex}.
The order parameter for superconductivity is the following hole-pair field,
\begin{equation}
M_{o\nu}=\bar{\psi}_{o\nu}(\bar{z}z_{o\nu})\eta_o,
\label{holepair}
\end{equation}
which has the electric charge $+2e$ and invariant under (\ref{gaugetrf}).


\section{Spin dynamics in chain and ladder }
\setcounter{equation}{0}

In this section, we study the spin part of $A_{\rm eff}$
by taking the continuum limit.
An essential difference appears between the ladder
and chain.

We shall first consider the case $J' \leq J$ and 
focus on the imaginary term in the effective model, i.e.,
the last two terms in (\ref{Az}).
\begin{eqnarray}
I^z &=& {1 \over 2J+J'} \sum_{o\nu}J_{\nu}\Big[ (\bar{\tilde{z}}\dot{z})
(\bar{z}z_{o\nu})(\bar{z}_{o\nu}\tilde{z}) +(\bar{z}\dot{\tilde{z}})
(\bar{\tilde{z}}z_{o\nu})(\bar{z}_{o\nu}z)\Big]  \nonumber   \\
&=& {1 \over 2J+J'} \sum_{o\nu}J_{\nu}\Big[(\bar{\tilde{z}}\dot{z})
(\bar{z}_{o\nu}\tilde{z})+(\bar{z}\dot{\tilde{z}})
(\bar{\tilde{z}}z_{o\nu})\Big]
+...
\label{Iz}
\end{eqnarray}
We introduce (continuous) coordinate $x=ai$ 
($a $ is  the lattice spacing).
For smooth configurations of $z$, we obtain
\begin{eqnarray}
\int d\tau I^z &\simeq &  {1 \over 2J+J'} a \sum_o (2J+J')\int d\tau
\Big( \overline{D_xz}D_{\tau}z
-\overline{D_{\tau}z}D_xz\Big)  \nonumber   \\
& \simeq & \int dxd\tau \Big( \overline{D_xz}D_{\tau}z
-\overline{D_{\tau}z}D_xz\Big)   \nonumber  \\
& \equiv & 2\pi i Q, 
\label{Tterm}
\end{eqnarray}
where $D_{\mu}=\partial_{\mu}+iA_{\mu}
=\partial_{\mu}-(\bar{z}\partial_{\mu}z)$,
and $Q$ is the topological charge which takes integer values for smooth
configurations of $z$ since it is a wrapping number from $S_2$ of 
the $x-\tau$
space  to $O(3)$ of the $\vec{S}$ space.
On the other hand, for a single quantum spin chain,
\begin{equation}
\int d \tau I^z_{chain}=\pi iQ.
\label{Tterm2}
\end{equation}
The reason why the coefficient of the topological term doubles in the 
ladder case is simply because there are twice as many degrees of freedom
per unit length in the ladder case.

Before going into the detailed study on the effect 
of the topological term,
let us obtain the continuum limit of the remaining terms 
in $A^z_0$ and $A^z_1$ in (\ref{Az}).
By straightforward calculation, we obtain
\begin{eqnarray}
A^z_0 & \simeq & -{1 \over 2}\sum_o\Big(J(2a)^2\overline{D_xz}D_xz+
J'(\sqrt{2}a)^2\overline{D_xz}D_xz \Big)  \nonumber   \\
& \simeq & -(2J+J')a\int dx\overline{D_xz}D_xz,  \nonumber   \\
A^z_1 & \simeq & {2 \over 2J+J'} 
\int dx\Big[-a^{-1}\overline{D_{\tau}z}D_{\tau}z
+(J+J'/\sqrt{2})^2a\overline{D_{x}z}D_{x}z\Big]+I^z.
\label{Azcon}
\end{eqnarray}
Therefore, the continuum limit of the spin part of $A_z \equiv A^z_0+A^z_1$
is the relativistic CP$^1$ model with the topological $\theta$-term.
\begin{eqnarray}
A_{CP} &=& {1 \over g^2} \int dx \sum_{\mu=\tau, x}
\overline{D_{\mu}z}D_{\mu}z +I^z,  \nonumber   \\
g^2 &=& {1+J'/2J \over \sqrt{1+(2-\sqrt{2})J'/J}}, 
\label{ACP}
\end{eqnarray}
where we have rescaled the imaginary time as $\tau \rightarrow v_z\tau$ and 
$v_z= [J(J+(2-\sqrt{2})J')]^{1/2} a$ is the 
``speed of light" of the present system.
Recently,  the $O(3)$  nonlinear-$\sigma$ model
with $O(3)$ variables $ \vec{n}(x)$ $(\vec{n}(x)\cdot\vec{n}(x) = 1)$  was derived
as an effective-low energy model of  the AF Heiseberg models on 
a ladder\cite{effaction}.
This field theory model has essentially the same structure
as (\ref{ACP}).
However, in order to discuss the hole-doped case, the use of CP$^1$ variables
is indispensable.

The nonlinear-$\sigma$ model with the $\theta$-term has been studied both 
analytically and numerically
as an effective field theory of the AF Heisenberg models\cite{CP1}.
Properties of  the ground state and excitations strongly depend on the 
value of $\theta$, i.e., the coefficient of the topological term,
$\int d\tau I^z(\theta)=i\theta Q$.    
It is expected that the model exhibits a 
phase transition at $\theta=(2m+1)\pi$ with $m=$ integer as $\theta$ varies.
Here we explain this transition from the view point of gauge theory.
It is well known\cite{IMCSS1} 
that a composite gauge field $A_{\mu}$ appears in the CP$^{N-1}$ model,
$A_{\mu}=i(\bar{z}\partial_{\mu}z)$, which transforms as $A_{\mu} \rightarrow
A_{\mu}-\partial_{\mu}\phi$ under $z \rightarrow e^{i\phi}z$.
For the case of $\theta=0$ (mod $2\pi$), the $z$-boson becomes massive
due to the CP$^1$ constraint \cite{const} 
and, after the $z$-integration,  the composite gauge boson acquires
the Maxwell term and so becomes dynamical.
In the $(1+1)$ dimensions, the Maxwell gauge theory has only one phase,
i.e., the confinement phase with a linear confining potential 
$V(r) \propto r $ between a pair of charged particles. Therefore
possible excitations in the CP$^1$ model
are the gauge-invariant triplet boson $\vec{n}=(\bar{z}\vec{\sigma}z)$.
These results are obtained for the CP$^{N-1}$ model by the 
$1/N$-expansion \cite{N}, but expected to be correct for the $N=2$ case also.

For $\theta=\pi$ (mod $2\pi$), it is expected that the 
behavior of the dynamical gauge field is quite different 
from the $\theta=0$ case.
In the present CP$^1$ model, the $\theta$-term is rewritten 
in terms of the gauge field (in the real-time formalism) as 
\begin{equation}
\int dt I^z(\theta)=i{\theta \over 2\pi} \int dxdt E,
\label{Itheta}
\end{equation}
where $E$ is the electric field.  

Coleman \cite{Coleman} gave the following semi-classical 
argument on the gauge dynamics of  QED$_2$ with the above $\theta$-term.
As $E=\partial_xA_0$, it is obvious that the effect 
of the $\theta$-term is interpreted 
as putting $\pm \theta /(2\pi)$ charges at the spatial infinities $x =
\pm \infty$.
In the case of $\theta=0$, an electric flux appears through the 
Gauss' law between a pair of  
oppositely charged sources (say, an electron and a posotron), 
giving rise to a  linear potential 
which confines an electron and a positorn
in one-spatial dimension.
When a pair of $\pm \theta/2\pi$ charges are put in the spatial infinities,
this confinement picture is not changed till the magnitude of charges
increases up to $\theta/2\pi=1/2$.
In the case $\theta/2\pi=1/2$, 
the ``ground state" supports an electric flux of magnitude $1/2$
lying along the entire space.
As a pair of electron and positron are put into the system at $x_1, x_2$, 
there appear  step-function-like jumps in the electric flux at $x_1,x_2$. 
The electric field is 
$E(x) = 1/2 $ for $x < x_1$ and $x_2 < x $ and $E(x) = 1/2 -1 = -1/2$,
but its magnitude is still $1/2$ for all 
points\cite{theta}.
Therefore, the energy, proportional to $E(x)^2$, 
does not change as $x_1, x_2$ are varied, so
there is {\em no} confining force between 
charges. Thus, at $\theta/2\pi=1/2$,
the system exhibits a phase transition from the confinement phase to
the deconfinement phase.
Similar behavior is expected also 
for the CP$^{N-1}$ model with the $\theta$-term.
The unsolved problem is whether the phase transiton 
at $\theta=\pi$ (mod $2\pi$) is of first
order or of second order.
It may depend on the magnitude of the coupling constant\cite{CP1}.
However, 
it is known \cite{massless} that the $S=1/2$ AF Heisenberg chain 
has gapless modes
and  spin-spin correlation functions have a power-law decay.
These low-energy properties are described by the $k=1$ Wess-Zumino-Witten model.
Therefore it is correct that the CP$^1$ model with a  $\theta$-term 
{\it that corresponds to } 
 the $S=1/2$ AF Heisenberg chain must have a second-order phase transition
 at $\theta=\pi$.
 
The above discussion on the composite gauge boson reveals the 
essential difference between the  spin chain and the spin ladder.
As we showed, one has $\theta=\pi$ for the chain,
 which leads to the deconfinement phase
of composite gauge boson, whereas one has $\theta=2\pi$ for the two-leg ladder, 
which leads to the  confinement phase.
We expect that also for the hole-doped case the $\theta$-term is ineffective
in the ladder t-J model and one may ignore its existence.
Detailed study of the dynamics of the composite gauge boson 
and quasi-excitations
in the ladder t-J model with doped holes will be given in the following section.

We have considered the case $J \geq J'$ so far.
Similar results are obtained for the case $J' \geq J$. 
The continuum limit of the spin part is again the CP$^1$ model
with $\theta=0$ in (\ref{ACP}), but the effective coupling constant
$g^2$ is given as \cite{g} 
\begin{equation}
g^2= \sqrt{{1 \over 2}+{J' \over 4J}},
\label{gcoupling}
\end{equation}
and $v_z=\sqrt{2J(J+{J' \over 2})}a$.
It is interesting to notice that for $\theta=0$ (mod $2\pi$)
the CP$^{N-1}$ model is asymptotically free and $g^2$ becomes large
at low energies.
This fact and Eqs.(\ref{ACP}) and (\ref{gcoupling}) 
suggest  $J'/J \rightarrow +\infty$ 
for the low-energy limit, as it is expected 
from the appearance of the energy gap for any finite value of $J' (\neq 0)$
in the ladder model. (The two points  $J'=0 $ and $J'=\infty $ may be
fixed points of renormalization group. )

The effect of doped holes on the dynamics of spins 
is examined by integrating out
the hole field in the effective model.
The  relevant terms come from $K+T_1$ in (\ref{Apsi}),  
\begin{equation}
\exp \Big[\int d\tau \Delta A^z\Big]=\int [D\psi][D\eta]
\exp\Big[\int d\tau (K+T_1)\Big].
\label{DAz}
\end{equation}
To this end, the hole-hopping expansion is quite useful and reliable
especially at sufficiently high $T$ \cite{YTIM}.
In the hopping expansion, the bare propagator of holes is obtained from
the $K$-term (\ref{Apsi}) as follows; 
\begin{eqnarray}
G_{\psi}(\tau_1-\tau_2) &=& \langle\psi_{i,a}(\tau_1) \bar{\psi}_{i,a}(\tau_2)\rangle \nonumber  \\
&=& {e^{-m(\tau_1-\tau_2)} \over 1+e^{-\beta m}}\Big[ \theta(\tau_1-\tau_2)
-e^{-\beta m}\theta(\tau_2-\tau_1)\Big],  \nonumber   \\
G_{\eta}(\tau_1-\tau_2) &=& \langle\eta_{i,a}(\tau_1) \bar{\eta}_{i,a}(\tau_2)\rangle \nonumber  \\ 
&=& -G^{\ast}_{\psi}(\tau_2-\tau_1),
\label{Ghop}
\end{eqnarray}
where we employ the temporal gauge $A_0=0$ for simplicity.
The hole concentration  is expressed as  
\begin{equation}
\langle\bar{\psi}_i \psi_i\rangle \equiv \delta ={e^{-\beta m} \over 1+e^{-\beta m}}.
\label{delta}
\end{equation}
 
From (\ref{DAz}) and (\ref{Ghop}), we obtain 
$\Delta A^z$  for high $T$ as follows;
\begin{eqnarray}
\int d\tau \Delta A^z &=& \int d\tau d\tau' \sum_{o,\nu}t_{\nu}^2b_{\nu}(\tau)c_{\nu}(\tau')
\cdot \langle \bar{\psi}_{o\nu}\bar{\eta}_o(\tau)\eta_o\psi_{o\nu}(\tau')\rangle  \nonumber  \\
& \sim & \delta(1-\delta) \beta \int d\tau \sum_{o\nu}t^2_{\nu}b_{\nu}(\tau)
c_{\nu}(\tau).
\label{DAz1}
\end{eqnarray}
It is straightforward to obtain the 
continuum limit of $\Delta A^z$ (\ref{DAz1}) as 
\begin{equation}
\Delta A^z=\int dx  \Big[ C_{\tau} a^{-1}\overline{D_{\tau}z}D_{\tau}z
+ C_xa\overline{D_xz}D_xz \Big],
\label{DAz2}
\end{equation} 
where
\begin{equation}
C_{\tau}={4(2t^2+t'^2)\beta \over (2J+J')^2},  \;\; 
C_x=2t^2\beta.
\label{Cs}
\end{equation}
Thus the effect of hole hoppings is incorporated in the form of 
renormalization of the relativistic CP$^1$ model 
but the $\theta$-term is {\em not} generated.
This renormalization increases the 
effective coupling $g^2$, hence increases the spin gap.
This is expected since  hoppings of  holes
should reduce the spin-spin correlations.
However, as $T$ is lowered, $\Delta A^z$ in (\ref{DAz2}) becomes to 
overestimate the hole-hopping effects, not only because one needs to
include higher-order effect of hopping expansion, but because
of correlations among holes.
As we explained before, there exists attractive force between holes 
sitting on the NN sites,
as $T_0$ term in (\ref{Apsi}) indicates.
This effective interaction between holes generates correlations 
among NN holes to  hinder the single-hole hoppings. In some
region, it may give rise to hole pairings and superconductivity
as discussed in Sect.6.

\section{Dynamics of composite gauge bosons and quasi-excitations
in the confinment phase}
\setcounter{equation}{0}

In this section we shall study the gauge dynamics of the t-J model
on a ladder. In Ref.\cite{IMCSS2} we studied the CSS of the t-J model
on a 2D lattice both in the slave-boson and SF representations, and
calculated the transition temperature $T_{\rm CSS}(\delta)$ of the 
confinement-deconfinement (CD) transition.
It was shown
that the CSS takes place at low $T$'s {\it below}  
$T_{\rm CSS}$, being compatible with experiments.
The method presented  in Sect.4 of Ref.\cite{IMCSS2}
can be applied in a straightforward manner 
to the study of the CSS in the present  ladder system.
Therefore we present below the main steps and results. 
The reader who wants to know more details should refer Ref.\cite{IMCSS2}.


The effective lattice model of Sect.4 treats even and odd sites
asymmetrically by integrating out the odd-site spins.
To make the calculations and presentation below 
 simpler  and more transparent, we introduce  
 another effective lattice model, which treats 
even and odd sites in a symmetric manner.
The new symmetric model   is very naturally obtained
from the asymmetric model by adding odd-site spin variables so that
it recovers the 
even-odd lattice symmetry.  
Both the previous asymmetric model and the 
present symmetric model have 
the same naive continuum limit in the spin part, i.e., 
the CP$^1$ non-linear sigma model.  
Since the relation between the asymmetric model  and the symmetric model 
are so intimate, these two models must fall into the same universality 
class; in particular the qualitative result on the CSS derived below should 
apply also to the asymmetric model.

To avoid confusion with the effective model in Sect.3 and to make the 
expressions more transparent, we change some  
notations; For example we use $x$ ( and $y$) to denote sites of the ladder. 
The partition function of the symmetric model  has the following 
path-integral representation;
\begin{eqnarray}
Z &=& \int [dz][d\zeta]exp(\int_{0}^{\beta}d\tau A),\nonumber\\
A&= &A_{z} + A_{\zeta} ,\nonumber\\
A_{z}& = & -\frac{1}{4\hat{J}}\sum_{x}|D_{\tau}z_{x}|^2
+\sum_{(xy)}\frac{J_{xy}}{2}
|\bar{z}_{y}z_{x}|^2 , \nonumber\\
A_{\zeta} & = & -\sum_{x}\bar{\zeta}_{x}(D_{\tau}+m_{x})\zeta_{x}
\nonumber\\
&+&\sum_{(xy)}
t_{xy}(b_{xy}\bar{\zeta}_{y}\bar{\zeta}_{x} + c_{xy}\zeta_{x}\zeta_{y}),
\label{eq:z}
\end{eqnarray}
where $\sum_{(xy)}$ denotes summation over NN pairs $(xy)$
on the ladder \cite{xy}. We write
$\hat{J}\equiv (2J + J')/4$, and $J_{xy}
\ (t_{xy})$
implies  $J (t)$ for vertical pairs and $J' (t')$ for horizontal pairs. 
$\zeta_{x}$ (Grassmann number) denotes
\begin{eqnarray}
\zeta_{x}=
\cases{
\psi_{x}& hole at even site,  \cr
\eta_{x} & anti-hole at odd site.}
\label{eq:zeta}
\end{eqnarray}
Due to this rewriting,
the fermion mass becomes staggered, i.e., $m_{x}= m $ for even sites
and $m_{x}= -m$ for odd sites.  

The spin part has been  simplified to the CP$^1$ lattice model by keeping
in $A_1^z$ only the first term.
The hole part is simplified by (i) ignoring $T_0$ and $T_2$
focusing on the region out of the
superconducting state, and (2) modifying $b_{\nu}$ and $c_{\nu}$ as
\begin{eqnarray}
b_{xy}&= &-\bar{\tilde{z}}_{x}{z}_{y}
+\frac{1}{2\hat{J}}(\bar{\tilde{z}}_{y}\dot{z}_{x}),
\nonumber\\
c_{xy}&=&-\bar{\tilde{z}}_{x}{z}_{y}+\frac{1}{2\hat{J}}
(\dot{\bar{z}}_{x}\tilde{z}_{y}).
\label{eq:bc}
\end{eqnarray}

The action is invariant under the local $U(1)$ gauge 
transformation
\begin{eqnarray}
z_{x\sigma} &\rightarrow&\exp(i\theta_{x})z_{x\sigma},\nonumber\\
\zeta_{x} &\rightarrow &\exp(i\theta_{x})\zeta_{x},
\label{eq:u1prime}
\end{eqnarray}
with time-independent rotation angles $\theta_x$. 


Next, we introduce the following four composite gauge variables 
on the link $(x,y)$; 
\begin{eqnarray}
B_{xy}\equiv &  (\bar{\tilde{z}}_{x}{z}_{y}) &\rightarrow  
e^{i\theta_{x}}B_{xy}e^{i\theta_{y}},\nonumber\\
D_{xy}\equiv & (\bar{z}_{x}{z}_{y}) & \rightarrow   
e^{-i\theta_x}D_{xy}e^{i\theta_{y}},\nonumber\\
F_{xy}\equiv & \zeta_{x}{\zeta}_{y} &  \rightarrow  
e^{i\theta_x}F_{xy}e^{i\theta_{y}},\nonumber\\
G_{xy} \equiv &\bar{\zeta}_{x}{\zeta}_{y}  &\rightarrow  
e^{-i\theta_x}G_{xy}e^{i\theta_{y}}
\label{eq:gauge}
\end{eqnarray}
where their transformation laws are also indicated.
We enforced these relations strictly via delta functions 
like $\delta(B_{xy} - \bar{\tilde{z}}_x z_y)$, etc.,
which  amounts to insert into $Z$ the following identity;
\begin{eqnarray}
1 &= &
\prod_{(xy)\tau}\int[dP][dQ][dR][dS]\int[dB][dD][dF][dG] \nonumber\\
 &&\exp \int d\tau \sum_{(xy)} i [P(B-(\bar{\tilde{z}}_{x}{z}_{y})) 
 +Q(D-(\bar{z}_{x}{z}_{y})) \nonumber\\   
 &+&R(F-\zeta_{x}{\zeta}_{y})
+S(G-\bar{\zeta}_{x}{\zeta}_{y} )+ h.c.]_{(xy)}.  
\label{eq:delta2}
\end{eqnarray}
Then by taking the temporal gauge, we have
\begin{eqnarray}
Z &=&\prod_{(xy) \tau} \int[dP][dQ][dR][dS]\int[dB][dD][dF][dG] 
\nonumber\\
&\times&I_{\zeta}(R,S) I_{z}(P,Q,F) \exp \int d\tau \sum_{(xy)} \tilde{A}_{xy}, 
\nonumber\\
I_{\zeta}(R,S)&=&\int [d\zeta]\exp\int d\tau[-\sum_{x}\bar{\zeta}_{x}(D_{\tau}
+m_{x})\zeta_{x}\nonumber\\
&-&i\sum_{(xy)}(R_{xy}\zeta_{x}{\zeta}_{y}+S_{xy}\bar{\zeta}_{x}{\zeta}_{y}+ h.c.)],\nonumber\\
I_{z}(P,Q,F) & = & \int [dz]\exp\int d\tau 
[-\frac{1}{4\hat{J}}\sum_{x}|\dot{z} _{x}|^2       
+\frac{1}{2\hat{J}}\sum_{(xy) }(t_{xy}\bar{F}_{xy}\bar{\tilde{z}}_{y}
\dot{z}_{x}-h.c.)
\nonumber\\
&-&i\sum_{(xy)}(-P_{xy} \bar{\tilde{z}}_{x}{z}_{y}
+Q_{xy}\bar{z}_{x}{z}_{y}+ h.c.)], \nonumber\\
\tilde{A}_{xy}&=&\frac{J_{xy}}{2}|D_{xy}|^2 
+t_{xy}(\bar{F}B + h.c.)_{xy}\nonumber\\
&+&i(PB+QD+RF+SG+h.c.)_{xy}
\label{eq:atilde}
\end{eqnarray}



Next, we integrate over $\zeta_x$ and $z_x$ by the hopping expansion;
an expansion in powers of gauge fields.
To do this we need the following on-site Green functions;
\begin{eqnarray}
\langle\zeta_{x}(\tau_{1})\bar{\zeta}_{y}(\tau_{2})\rangle& = &\delta_{xy}
 G_x(\tau_1 - \tau_2),\nonumber\\
G_x(\tau)& =& \frac{1}{\beta} \sum_{n= -\infty}^{\infty} 
\frac{\exp(i\omega_{n} \tau)}{i\omega_n + m_x}\nonumber\\
& = & \mbox{sgn}(m_x)\sum_{L} (-)^L \exp[-m_x(\tau+\beta L)]
\theta[m_x(\tau+\beta L)]
\nonumber\\
&=& \frac{\exp(-m_x \tau)}{1+\exp(-m_x \beta)} 
[ \theta(\tau)-\theta(-\tau)\exp(-m_x \beta )],   
\label{eq:g} 
\end{eqnarray}
for $|\tau | <\beta$, and
\begin{eqnarray}
\langle z_{x\sigma}(\tau_{1})\bar{z}_{y {\sigma}^{'}}(\tau_{2})\rangle & = &\delta_{xy}
\delta_{\sigma {\sigma}^{'}}G(\tau_1 - \tau_2),\nonumber\\
G(\tau)& =& \frac{4\hat{J}}{\beta} \sum_{n= -\infty}^{\infty}
 \frac{\exp(i\omega_{n} \tau)}{\omega_n^2 + \sigma} \;\;\;
( \omega_{n} \equiv 2\pi n/\beta) \nonumber\\
& = & \frac{2\hat{J}}{\sqrt{\sigma}}\sum_{K} 
\exp[-\sqrt{\sigma}(\tau+\beta K)]
\nonumber\\
&=& \cases{
\frac{4\hat{J}}{\sigma}\delta(\tau) & for $\sqrt{\sigma}\beta >> 1$ \cr
\frac{4\hat{J}}{\beta\sigma}& for $\sqrt{\sigma}\beta << 1 $
}
\label{eq:gz}
\end{eqnarray}
Value of $\sigma$ can be estimated in the 
large-$N$ approximation, etc. \cite{IMCSS1}, but it is not necessary to
derive the main conclusion below as long as it is finite. 

Up to the second order in the  gauge variables, we obtain 
\begin{eqnarray}
Z & = & \int [dB][dD][dF][dG][dP][dQ][dR][dS]
\exp\int d\tau\sum_{xy}[A_{\zeta z}+
\nonumber\\
& + & \frac{J_{xy}}{2}|D|^2
 +t_{xy}(\bar{F}B + h.c.)\nonumber\\
&+&i(PB+QD+RF+SG+H.c.)]_{xy},\nonumber\\
A_{\zeta z}& = & -c_1 \beta|R_0|^2-c_2 |S|^2 +c_3 |\dot{S}|^2-c_4 |F|^2 
-c_5 |\dot{F}|^2-c_6 |P|^2 \nonumber\\
&+&c_7 |\dot{P}|^2-c_8 |Q|^2 +c_9|\dot{Q}|^2-i c_{10}(\bar{F}
\dot{\bar{P}}- h.c.)
\label{zz2}
\end{eqnarray}
where the coefficients $c_{1,2,...,10}$ are given by 
\begin{eqnarray}
c_1 & = & g\;\exp(-\beta|m|),\;\;\;\;\;\; g\equiv (1+\exp(-\beta|m|)^{-2},
\nonumber\\
c_2& = & \frac{g}{2|m|},\;\; c_3 =  \frac{g}{8|m|^3},
 \;\;c_4 =  \frac{2t_{xy}^2}{\sigma^{1/2}}, \;\; 
 c_5 = \frac{t_{xy}^2}{2\sigma^{3/2}},
\nonumber\\
c_6 & = & c_8= \frac{8\hat{J}^2}{\sigma^{3/2}},\;\; 
c_7 = c_9= \frac{2\hat{J}^2}{\sigma^{5/2}},\;\;
c_{10} = \frac{t_{xy}\hat{J}}{\sigma^{3/2}} 
\label{eq:c}
\end{eqnarray}
$R_0$ is the $n=0$ mode of the Fourier coefficients, $R_n \equiv
\beta^{-1}\int_0^{\beta}  d\tau R(\tau) exp(-i\omega_n \tau)$. 


The integrations over $P,Q,S$ are Gaussian, and 
the $R(\tau)$-integration is done in  Fourier components
$R_n$;  
\begin{eqnarray}
&&\int \prod_n dR_n  \exp[-c_1 \beta^2 |R_0|^2+i\beta 
\sum_n(R_n F_{-n} + h.c.)] 
\nonumber\\
&\propto&\exp(-\frac{1}{c_1} |F_0|^2)
\prod_{n \neq 0} \delta(F_n)
\label{eq:r}
\end{eqnarray}

We further integrate over the pair fields $F$ and $G$ of fermions.
To discuss the phase dynamics of residual gauge variables $B$ and $D$,
we set their amplitudes to be the constants $\rho_{B,q}(\equiv|B_{xy}|)$
 and  
$ \rho_{D,q}(\equiv|D_{xy}|)$, which can be determined by a straightforward
mean field theory ignoring phase fluctuations of mean fields. The suffix $q$ =(u,s) distinguishes
whether the link $(x,y)$ is vertical, $(x(i,a),y(i+1,a))$  or horizontal $(x(i,1),y(i,2))$\cite{xy}. 
 Thus we write 
\begin{eqnarray}
B_{xy} &=&\rho_{B,q}U_{xy}, U_{xy} \in U(1),\nonumber\\
D_{xy}& =&\rho_{D,q}V_{xy}, V_{xy} \in U(1).
\label{eq:uv}
\end{eqnarray}
Due to the compliteness condition, 
$z_{\sigma}\bar{z}_{\sigma^{'}}+\tilde{z}_{\sigma}
\bar{\tilde{z}}_{\sigma^{'}}
=\delta_{\sigma \sigma^{'}}$, we have the relation,
\begin{eqnarray}
\rho_{B,q}^2 + \rho_{D,q}^2 = 1.
\label{eq:complete}
\end{eqnarray}
After some  calculations, we reach the following 
effective lattice gauge theory;
\begin{eqnarray}
Z & = & \int [dU][dV] \exp(\int d\tau \sum_{(xy)}A_{UV})\;,\nonumber\\
 A_{UV}& =&
-\hat{c}_{D,q}  |\dot{U}_{xy}|^2 
-\hat{c}_{B,q}  |\dot{V}_{xy}|^2 \nonumber\\
\hat{c}_{D,q} & = & (\frac{c_9}{c_8^2})\rho_{D,q}^2, 
\hat{c}_{B,q} =  [\frac{c_7}{c_6^2  } 
+ \frac{\beta^3}
{(2\pi)^2}\frac{c_1 t_{xy}^2}{1 + \beta c_1 c_4}] \rho_{B,q}^2, 
\label{eq:zu}
\end{eqnarray}
\par

Following the Polyakov-Susskind method \cite{PS} 
for studying CD transitions in lattice gauge theories, 
this system can be mapped 
to the anisotropic XY model\cite{IMCSS2};
\begin{eqnarray}
Z_{XY} &= & \int \prod_{x} \frac{d \alpha_{x}}{2\pi} \exp
[\sum_{(xy)}J_{1,q} \cos (\alpha_{y}-\alpha_x) +\sum_{(xy)}J_{2,q} \cos 
(\alpha_{y} + \alpha_x)], \nonumber\\
  J_{1,q} &\equiv  & \beta^{-1} \hat{c}_{D,q}, \ 
  J_{2,q} \equiv   \beta^{-1} \hat{c}_{B,q}.
\label{eq:dXY4}
\end{eqnarray}
The $J_2$ term, coming from $|\dot{V}|^2$ term, 
 expresses an anisotropy of XY spin couplings, and reduces the global 
 spin symmetry from $U(1)$ (for  $J_2 =0$) down to $Z(2)$
 (for  $J_2 \neq 0$).
A possible phase transition of this spin system describes the
CD phase transition of the original t-J model.
For two and higher dimensional lattices, this spin system is known 
to have an order-disorder transition at $J_1 + J_2 \simeq 1$
\cite{IMCSS2},  which leads to existence of a finite 
transition temperature $T_{\rm CSS}$.
For $T > T_{CSS}$, the XY spin correlations decay exponentially,
$ f(|x|) \equiv \langle\vec{S}_x \vec{S}_0\rangle \simeq \exp(-a |x|)$, which implies
the potential energy among two gauge charges, $W(|x|) = -\ln f(|x|)$
(see Ref.\cite{PS} for this relation), to be 
a linear-rising confining potential, $W(|x|) \simeq a |x|$. So
the gauge dynamics  realizes here in the confining phase. For $T < T_{CSS}$,
there is a long-range order, $f(|x|) = const + \exp(-c|x|)$, which
implies a short-range potential $W(|x|) \simeq \exp(-c|x|)$.
So the gauge dynamics  realizes in the  deconfining  
phase here\cite{KT}.

In the present ladder system, the XY model (\ref{eq:dXY4})
is put on a coupled two one-dimensional chains and
its universality class is  classified into that of a single 
one-dimensional chain,
the partition function of which is given by
\begin{eqnarray}
Z_{XY}^{\rm chain} &= & \int \prod_{i} \frac{d \alpha_{i}}{2\pi} \exp
[J_1\sum_{i}\cos (\alpha_{i+1}-\alpha_i) +J_2\sum_{i} \cos 
(\alpha_{i+1} + \alpha_i)].
\label{eq:dXY5}
\end{eqnarray}
One can solve its spin correlations exactly, which decay 
exponentially, $f (|x|) = A\exp(-a|x|)$. The coefficient $a$ is finite 
as long as $J_{1,2}$ are finite. Thus, we conclude that
the gauge dynamics of the t-J ladder is always realized in 
the confining phase except for $T= 0$ at which $J_{1,2}$ diverge.
This is the main conclusion of this section.
The case of $T=0$ is discussed later on.
 
Here we comment that the above conclusion is based on the hopping expansion.
If $\sigma$, the mass of $z$, vanishes, the hopping expansion 
cannot be applicable. Our assumption $ \sigma > 0$ is supported
by the large-$N$ analysis of the CP$^{N-1} \sigma$ model in one spatial
dimension \cite{IMCSS1}.
This is in strong contrast with the case of  a single chain.
There, the topological term appears and it generates massless excitations,
preventing us from making use of the hopping expansion and obtaining
 an effective lattice gauge theory. However, 
as discussed in Sect.4, we know that the spin chain system is realized 
in the deconfinement phase. We expect that the doped chain
still supports this deconfinement phase, although a solid
analysis is required.
(The special case of supersymmetry $J/t=2$ can be solved exactly by the
Bethe anzatz and its result supports the CSS.)

How about the gauge dynamics on a ladder at  $T=0$ ? Since the system
(\ref{eq:dXY5}) develops a long-range order,  one may conclude
that the gauge system is in the deconfinement phase. However,
the expression  (\ref{eq:dXY5}) itself is not adequate at $T=0$.
This can bee seen as follows. At $ T = 0$, $m =0$ as one can see from 
(\ref{delta}). So the hopping expansion with respect to fermions
is not applicable. In this case, we have a separate argument using 
the massless Schwinger model \cite{Schwinger}, 
the massless QED$_2$ without topological
terms, which is obviously
a relevant model for present argument.  Here it is known \cite{Schwinger} that
the physical spectrum consists of  a massive   boson, a composite of 
fermion and antifermion.  Since this is a gauge-invariant neutral object,
its gauge  dynamics is  in the confinement phase.
From this fact, we  conclude  that the t-J model on a ladder realizes its
gauge dynamics in the confinement phase not only for $T > 0$ but also
for $T=0$.

How is the effect of weak but finite couplings among NN ladders in a plane
and couplings among interlayers ?  They put the system into 
three dimensional one. The calculation given above can be redone
to reach the same expression  (\ref{eq:dXY4}) for the XY spin  model, 
but the sum over $(xy)$ is extended to the entire lattice. One concludes
that   there is a finite critical temperature $T_{CSS}$ {\it below} 
which the deconfinement phase appears. The value $T_{CSS}$
is small and approaches zero as these three-dimensional $J$ and $t$ 
couplings tend to vanish. One can determine $T_{CSS}$ 
using the 3D XY model and the MF values of $\rho_{B,q}$, etc. 

What are the quasi-excitations in the confinement phase ?
In this phase, due to the confining force,  only charge-neutral objects 
can be physical excitations. 
The typical examples are the original electrons
$\bar{z}_x\psi_x$ which carry  the electric charge $-e$ and spin $S = 1/2$,
the charge density fluctuation $\bar{\psi}_x\psi_x$, 
and the spin triplet excitation
$\bar{z}_x \vec{\sigma} z_x$, i.e., charge-neutral magnons of $S=1$. 
Together with these local combinations, one may conceive nonlocal 
gauge-invariant combinations like $\bar{z}_x U_{xy}\psi_y$, etc.
Real low-energy excitations are linear combinations of these states 
with definite electric charge and spin.

Finally, we note that the analysis in this section can be repeated
when one includes the $T_0$ and $T_2$ terms into the symmetric model.
These terms are necessary to generate the superconducting phase
as we shall see in the following section.
The corresponding calculations of $T_{CSS}$  for the 2D t-J model 
are found in Ref.\cite{IMCSS2}. After tracing these calculations, one 
reaches the same effective XY spin model of 
(\ref{eq:dXY4}) with the modified coefficients $J_{1,2}$.
Thus we obtain the conclusion that the confinement phase 
obtained above continues to exist even when the system is superconducting.
  
\section{Superconductivity}
\setcounter{equation}{0}

As we saw in Sect.4, there are effective interactions
between holes which appear as a result of SRAFO.
In terms of the composite hole-pair field $M_{o\nu}$, 
$T_0$ and $T_2$ terms in (\ref{Apsi}) 
are rewritten as 
\begin{eqnarray}
T_0 &=& {1 \over 2}\sum_{o,\nu} J_{\nu}\bar{M}_{o\nu}M_{o\nu},  \nonumber  \\
T_2 &=&-{2 \over 2J+J'}\sum_o\sum_{\nu\nu'}t_{\nu}t_{\nu'}\bar{M}_{o\nu}M_{o\nu'}.
\label{T0T2}
\end{eqnarray}
It is obvious that $T_0$ induces a condensation of the hole-pairs,
\begin{equation}
\langle M_{o\nu}\rangle  \neq 0,
\label{dwave}
\end{equation}
and $T_2$ favors the ``d-wave" symmerty,
\begin{equation}
\sum_{\nu}t_{\nu}\langle M_{o\nu}\rangle=0.
\label{dwave2}
\end{equation}
For $t'=t$, Eq.(\ref{dwave2}) 
means $\langle M_{os}\rangle=-2\langle M_{ou}\rangle
=-2\langle M_{od}\rangle$ if we assume $\langle M_{o,u }\rangle = \langle M_{o,d}\rangle$ and $\langle M_{o,s(u,d)}\rangle$ 
  are independent of the position $o$ \cite{flux}.
 
It is straightforward to introduce a hole-pair field 
$\Delta_{o\nu}$ by a Hubbard-Stratonovich  transformation.
We shall obtain a Ginzburg-Landau (GL) theory of the hole-pair field
by integrating out the hole field.
To this end we take the temporal gauge $A_0=0$ for calculational 
simplicity\cite{full}.
The $T_0$ term is rewritten by $\Delta_{o\nu}$ as
\begin{equation}
\exp (\int d\tau T_0) =\exp \Big(\int d\tau 
\sum_{o\nu}\frac{J_{\nu}}{2}(M_{o\nu}
\bar{\Delta}_{o\nu}+\bar{M}_{o\nu}\Delta_{o\nu}
- \bar{\Delta}_{o\nu}\Delta_{o\nu})\Big),
\label{Dhole}
\end{equation}
up to an irrelevant constant. From this, one gets the relation,
\begin{eqnarray}
\langle M_{o\nu} \rangle &=& \Delta_{o\nu}.
\end{eqnarray}
The GL potential energy, $V_{\rm GL}(\Delta)$ of $\Delta_{o\nu}$ (that appears in 
the action integral $\int d\tau A$   in the form of $-\beta V_{\rm GL}$ ), is obtained by the hole-hopping expansion as 
\begin{eqnarray}
V_{\rm GL}(\Delta) &=&  V_0+V_1+V_2 +O(\Delta^4),   \nonumber   \\
V_0&=&  {1 \over 2} \sum_{o\nu}J_{\nu}\bar{\Delta}_{o\nu}\Delta_{o\nu}, \nonumber \\
V_1 &=& -\frac{1}{\beta}\Big({1 \over 2}\Big)^2 \int d\tau d\tau' \sum_{o\nu}J^2_{\nu}
\bar{\Delta}_{o\nu}\Delta_{o\nu}\cdot \langle M_{o\nu}(\tau)\bar{M}_{o\nu}
(\tau')\rangle,  \nonumber   \\
V_2 &=& \frac{1}{\beta} \Big({1 \over 2}\Big)^2 {1 \over J+J'/2}
\int d\tau d\tau' d\tau''\sum_{o\nu}J_{\nu}J_{\nu'}
t_{\nu}t_{\nu'}\bar{\Delta}_{o\nu}\Delta_{o\nu}   \nonumber  \\
&& \;\;\times\langle \bar{M}_{o\nu}(\tau)
M_{o\nu'}(\tau)M_{o\nu}(\tau')\bar{M}_{o\nu'}(\tau'')\rangle,
\label{V012}
\end{eqnarray}
where $\Delta_{o\nu}$ are assumed to be time-independent.
The above correlation functions of $M_{o\nu}$'s are calculated by the
hole propagator (\ref{Ghop})
and the expectation value  of $z$-pair, 
\begin{equation}
\langle  \bar{z}z_{o\nu}  \rangle = D_{\nu}.
\label{Dz}
\end{equation}
The actual value of $D_{\nu}$ should be obtained by the MFT\cite{IMS},
 but here we only needs that
 it is nonvanishing due to SRAFO.
After some calculation, we obtain  
\begin{eqnarray}
V_1 &=&   {1 \over 8m}\sum_{o\nu}
 J_{\nu}^2 |D_{\nu}|^2 \bar{\Delta}_{o\nu}\Delta_{o\nu}, \nonumber  \\
V_2 &=&- {1 \over J+{J' \over 2}}  {1 \over 16m^2} \sum_o 
\Big(\sum_{\nu}t_{\nu}J_{\nu}D_{\nu}\Delta_{o\nu}\Big)
\Big(\sum_{\nu'}t_{\nu'}J_{\nu'}\bar{D}_{\nu'} \bar{\Delta}_{o\nu'}\Big).
\label{V12}
\end{eqnarray}
We are interested in the  case of low hole concentrations $\delta \ll 1$ where
we have the  expression, 
\begin{equation}
m \simeq \frac{|\ln \delta|}{\beta},
\label{mdelta}
\end{equation}
which comes from  (\ref{delta}). Since $ m$ is large for $\delta \ll 1$,
we have neglected the terms of $O(m^{-3})$ in the above.

Let us find a solution that minimizes $V_{\rm GL}(\Delta)$.
It is easily found by minimizing $V_0 + V_1$ and $V_2 $ separately.
By writing  $\Delta_{o\nu} = e_{\nu} \Delta$ , we have
\begin{eqnarray}
V_0 + V_1  & = & N_{\rm link} \times  C_2 |\Delta |^2 \nonumber\\
C_2 & = & \sum_{\nu}( -\frac{1}{2}J_{\nu}  + \frac{1}{8m}|D_{\nu}|^2
J_{\nu}^2) e_{\nu}^2. 
\label{V0+V1}
\end{eqnarray}
$V_2$ is minimized by  configurations satisfying
\begin{eqnarray}
&& \sum_{\nu} t_{\nu}J_{\nu}D_{\nu}\Delta_{\nu} = 0.
\label{V2sol}
\end{eqnarray}
Eq.(\ref{V0+V1}) implies that the order parameter $\Delta$ devevlops
for  $C_2  < 0$. So  the critical temperature $T_c$ is calculated
 from $C_2 = 0$ as
\begin{equation}
T_c = \frac{\sum_{\nu}J_{\nu}^2 D_{\nu}^2 e_{\nu}^2}
{\sum_{\nu}J_{\nu} e_{\nu}^2} \frac{1}{4 |\ln \delta|} \sim
\frac{D^2 J }{4 |\ln \delta|}.
\label{Tc}
\end{equation}
The last expression is for the case of $J'=J$, $t'=t$ and $\nu$-independent
$e_{\nu}$ and $D_{\nu} = D$. 
In this case,    
Eq.({\ref{V2sol}) gives rise to the $d$-wave solution \cite{flux},
\begin{equation}
\Delta_{os}=-2\Delta_{ou}=-2\Delta_{od}.
\label{dwave3}
\end{equation}
The MFT using a  Gutzwiller renormalization 
of matrix elements\cite{supth} predicts
similar behavior of the superconductivity correlations.
If we take into account the fluctuations of the order parameter of the
superconductivity $\Delta_{o\nu}$, weak but finite three-dimensional
inter-ladder interactions are required to realize a genuine
superconducting phase. 
From our discussion given so far, it is obvious that the existence of the
SRAFO is essential for appearance of the superconductivity and
its d-wave type symmetry.

\section{Discussion}
\setcounter{equation}{0}  
 
In this paper, we studied the t-J model on the ladder by using the 
SF-CP$^1$ formalism.
We obtained the effective low-energy model by integrating out the half of the
CP$^1$ variables.
At low $T$'s, there exists the SRAFO which makes it easy
to integrate over the CP$^1$ variables.
The low-energy effective model shows why the low-energy excitations
are different in the ladder and  the chain systems, reflecting the 
coefficient of the topological term in the CP$^1$ model.
In the ladder system, the confinement mechanism works at any $T$
and therefore excitations are spin triplet and spinon-holon bound state.
On the other hand, in the chain case, the full CSS takes place because
of the $\theta$-term of the composite gauge boson.
Similar analysis works for the 2D t-J model, 
which is studied in the previous papers.
We stressed that the gauge-theoretical point of view,
especially the CD transition, is quite
useful and universally applicable for such problems of
separation of degrees of freedom.

For even-number-leg ladder Heisenberg models, it is expected that
an energy gap appears for spin excitations for any finite value of $J'$.
The continuum field theory model is the CP$^1$ model which is known
to be  asymptotically  free.
Therefore the coupling constant becomes large at low energies.
Our derivation shows that this means $J'/J \rightarrow +\infty$ for
low-energy limit as expected.
    
The effective model also shows that there are effective 
attractive interactions
between holes in the SRAFO background.
By these interactions, the superconducting phase is possible 
under weak but finite 3D inter-ladder interactions.

In conclusion, we observed that the SF-CP$^1$ formalism and the 
gauge-theortical method provide a natural and coherent way to
understand various important properties of
the t-J model on a ladder.


\eject

Figure Caption\\

Fig.1 Layout of a ladder. Two vertical lines denotes two legs, $a = 1$ 
and $ a = 2$. A site on the ladder is labelled by a set of suffices $(i,a)$,
where $i$ counts the site along each leg. The filled circles denote
odd sites, and the open circles denote even sites.
For a fixed odd site $o$, there are three nearest neighbor 
even sites, $(o,\nu)$ with
$\nu = u$  (up), $\nu = d$ (down), $\nu = s$ (side). For example, 
for $o =  (i,1) , \ (o,u) = (i+1,1), \ (o,d) = (i-1,1), \ (o,s) = (i,2)$ as illustrated in the Figure. For $o =  (j,2), \ (o,u) = (j+1,2), \ (o,d) = 
(j-1,2), \ (o,s) = ( j,1)$. The hopping amplitude and the exchange coupling are
$t,J$, respectively,  for vertical links $(i,a),(i + 1,a)$, and $t',J'$ for horizontal links $(i,1),(i,2)$.


\begin{thebibliography}{1}
\bibitem{const}
The constraint can be respected 
by adding the term $\int dx 
\lambda(x)(\bar{z}(x) z(x) -1)$
in the action, where $\lambda(x) $ is the Lagrange multiplier field.
For examaple, in the large-$N$ analysis, $\lambda(x)$ developes a 
nonvanishing expactation values $\lambda_0$ 
and fluctuations around it is suppressed by $1/N$. Thus this term
reduces to the mass term of $z$ field with the mass square $\lambda_0$
\bibitem{theta}
When $|\theta|/2\pi >1$,
spontaneous generation of electron-positron pair occurs to shield
charges at infinities.
Therefore the system must be periodic with respect to $\theta$ mod $2\pi$.
\bibitem{xy}
For definiteness,  we assign $y(i,a)$ for given $x(i,a)$ as follows;
$y(i+1,1)\  {\rm or }\  y(i,2) \    {\rm for }\  x(i,1)$ and
$y(i+1,2) \  {\rm for } \ x(i,2)$.
\bibitem{KT}
For $J_2 = 0$ in 2D lattice, the transition is
the Kosterlitz-Thouless transition, and $f(|x|) \simeq |x|^{-\eta},
W(|x|) \simeq ln(|x|)$. This is a quasi-confining phase.
\bibitem{full}
A full gauge-theoretical study of 
the composite gauge bosons
is of course possible.
Details of that will be reported in a future publication.
\end{thebibliography}
 \end{document}